\documentclass[runningheads,natbib]{svjour2}

\usepackage{graphicx}

\journalname{Solar Physics}

\begin{document}

\title{ The Depiction of Coronal Structure in White Light Images}
\titlerunning{ Depiction of Coronal Structure}

\author{Huw Morgan \and
	 Shadia Rifai Habbal \and
	 Richard Woo}
	 
\institute{H. Morgan and S.R. Habbal \at 
Institute for Astronomy, University of Hawaii, 2680 Woodlawn Drive, Honolulu, HI 96822, USA \\
Tel.: 808-956-8319\\
\email{hmorgan@ifa.hawaii.edu}
\and
R. Woo \at Jet Propulsion Laboratory, 4800 Oak Grove Drive, MS 238-725, Pasadena, CA 91109, USA}

\date{}

\authorrunning{Morgan, Habbal and Woo}

\maketitle

\begin{abstract}
The very sharp decrease of density with heliocentric distance makes imaging of coronal density structures out to a few solar radii challenging. The radial gradient in brightness can be reduced using numerous image processing techniques, thus quantitative data are manipulated to provide qualitative images. Introduced in this study is a new normalizing radial graded filter (NRGF), a simple filter for removing the radial gradient to reveal coronal structure. Applied to polarized brightness observations of the corona, the NRGF produces images which are striking in their detail. Total brightness white light images include contributions from the F corona, stray light and other instrumental contributions which need to be removed as effectively as possible to properly reveal the electron corona structure. A new procedure for subtracting this background from LASCO C2 white light total brightness images is introduced. The background is created from the unpolarized component of total brightness images and is found to be remarkably time-invariant, remaining virtually unchanged over the solar cycle. By direct comparison with polarized brightness data, we show that the new background subtracting procedure is superior in depicting coronal structure accurately, particularly when used in conjunction with the NRGF. The effectiveness of the procedures is demonstrated on a series of LASCO C2 observations of a coronal mass ejection (CME). 
\end{abstract}

\section{Introduction}
\label{intro}

White light and polarized brightness ($pB$) observations have long been used to determine the density and large scale topology of the solar corona \citep[for example]{van1950,guh1996,que2002}. However, unprocessed images of the extended corona are not useful as qualitative indicators of coronal structure since they are dominated by the sharp gradient in brightness with increasing coronal height (radial gradient). The K corona brightness decreases by a factor of $\sim$10$^{4}$ starting near the solar limb and out to $3R_\odot$ \citep{hie2000}. Many techniques have been developed to lessen or remove the radial gradient. Although unsuitable for quantitative analysis, the general shape and distribution of streamers and coronal holes in processed images have strongly influenced our basic concepts of the topology of the coronal magnetic field \citep{woo2005}. Furthermore, white light images continue to be used to place observations by other remote sensing observations as well as in situ measurements of the solar wind distant from the Sun in their large scale coronal context \citep[for example]{gos1981}. The results of coronal simulations or models such as potential source surface models are also often compared with these processed white light images \citep[for example]{lie2001}. It is important therefore that processed white light images provide a true picture of coronal structure.

There are many techniques to overcome the sharp radial gradient in brightness. These we call radial graded filters (RGF).  Hardware RGF can be implemented at the time of observation using mechanical \citep{owa1967} or optical \citep[for example]{new1968} means. Alternatively, a coronal image can be compiled from a sequence of photographs taken with a variety of exposure times. The long exposures reveal the faint structure, and the short exposures reveal the bright structures without saturation. Such a procedure is described in \citet{gui1999}. These techniques allow the detector medium (photographic film or digital CCD) to record the large dynamic range of the white light corona. In the last few decades, digital processing of coronal white light images has become common practice. Fine scale coronal structure can be revealed with techniques such as unmasking or edge enhancing filters \citep{kou1988,kou1992} or even with standard photo-editing software \citep{esp2000}. 

This work is concerned with the digital processing of coronagraph data to form images which represent the large scale structure of the corona. Section \ref{nrgfsec} introduces a normalizing radial graded filter which removes radial gradient exactly, revealing excellent detail in observations. Section \ref{whitelight} describes a novel technique for removing background from Large Angle and Spectrometric Coronagraph (LASCO) C2 total brightness white light observations. The power of these techniques is demonstrated in section \ref{cme} on a series of consecutive observations of a propagating coronal mass ejection (CME). A short discussion is given in section \ref{conclusions}.

\section{The Normalizing Radial Graded Filter}
\label{nrgfsec}

\begin{figure}
\centering
\includegraphics[width=0.9\textwidth]{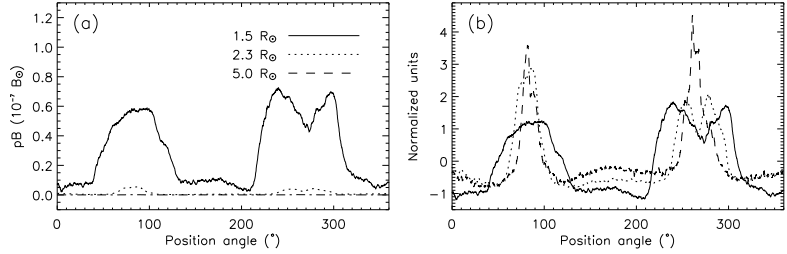}
		\caption{(a) Latitudinal profiles of pB extracted at three heights for the solar maximum corona of 1997/01/18. The profile at the lowest height is taken from the daily average of observations by the Mauna Loa Solar Observatory's (MLSO) MKIII coronameter and the other two from LASCO C2 observations. (b) The profiles shown in the top left plot normalized by subtracting their mean and dividing by their standard deviation.}
\label{f1}
\end{figure}

Figure \ref{f1}(a) illustrates the basic challenge of viewing structure in the extended corona. Three latitudinal profiles of pB are shown at heights of 1.5, 2.3 and 5.0$R_\odot$. It is of course almost impossible to compare structure in this linear plot due to the steep drop in brightness. For the same reason 2D coronal images without lessening of the radial gradient are not useful as qualitative indicators of coronal structure. In \ref{f1}(b), each latitudinal profile is normalized to a mean of zero and a standard deviation of one. This is a far better way to compare structure at all heights although all quantitative information is lost. We can generalize this normalizing approach to two dimensional images. In this case, we call the technique a normalizing radial graded filter (NRGF). Processed image intensity is given by 

\begin{equation}
I^{\prime}(r,\phi) = [I(r,\phi) - I(r)_{<\phi>}]/ \sigma(r)_{<\phi>},
\label{nrgf}
\end{equation}

\noindent where $I^{\prime}(r,\phi)$ is the processed and $I(r,\phi)$ is the original intensity at height $r$ and position angle $\phi$. $I(r)_{<\phi>}$ and $\sigma(r)_{<\phi>}$ are the mean and standard deviation of intensities calculated over all position angles at height $r$. 

\begin{figure}
\centering
\includegraphics[width=0.75\textwidth]{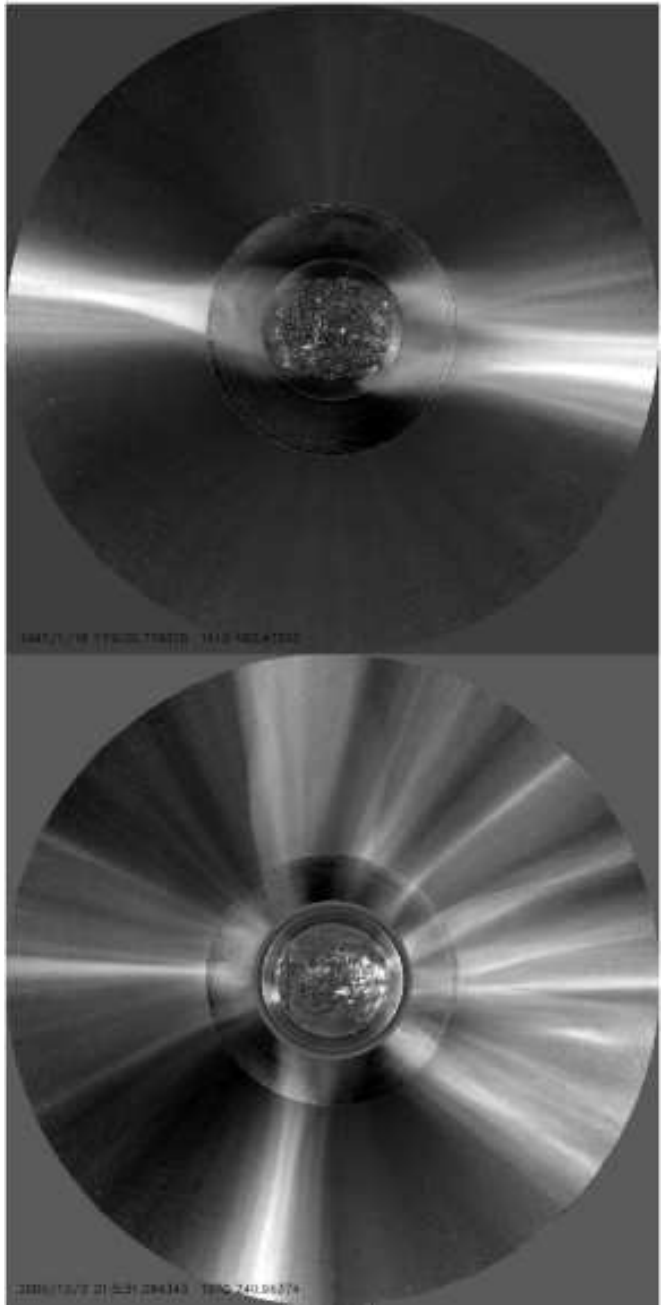}
		\caption{Composite coronal images processed by the NRGF at all heights above the Sun's limb for 1997/01/18 at solar minimum (top) and 2000/12/03 at solar maximum (bottom). The innermost views of the disk and low corona are EIT images of the He II 304\AA\ line. The corona from 1.15 to 2.3$R_\odot$ are from MLSO MKIII (minimum) and MKIV (maximum) observations and the outer fields of view from 2.3 to 6$R_\odot$ are from LASCO C2 observations.}
\label{f2}
\end{figure}

The results of applying the NRGF to images of the solar minimum and maximum corona are shown in figure \ref{f2}. These images are composed of observations in He II 304\AA\ by the Extreme Ultraviolet Imaging Telescope (EIT), and pB observations by the Mauna Loa Solar Observatory's (MLSO) MKIII (solar minimum) and MKIV (solar maximum) coronameters and the LASCO C2 coronagraph. The EIT images are cropped to a maximum height range of 1.3$R_\odot$. The MLSO coronameters observe polarized brightness in white light at heliocentric distances of $\sim$1.1 to 2.4$R_\odot$. Details of the MKIII instrument are given in \citet{fis1981}. The MLSO images are cropped to a height range of 1.3-2.3$R_\odot$. The LASCO instrument aboard SOHO has a set of three coronagraphs, each optimized to view different heights in the corona \citep{bru1995}. We use here $pB$ observations made by the C2 coronagraph, which has a height range of $\sim$2 to 6$R_\odot$. We crop the C2 images to a height range of 2.3-6.0$R_\odot$. After cropping, the data from the 3 instruments are combined, and a 2D image is formed using linear triangulation. The mean and standard deviation of $pB$ as a function of height is calculated and each image pixel is transformed using equation \ref{nrgf}. 

As can be seen in figure \ref{f2}, a useful aspect of the NRGF is the ease it allows to join images from various instruments. EIT measures emission in HeII 304\AA\, which is a very different  measurement to pB. Also there is usually a discontinuity between the MLSO and LASCO C2 measurements particularly in low signal regions. The simple normalizing function of equation \ref{nrgf} forgives these differences and results in excellent continuity at the boundaries between the instruments field of views without any smoothing or interpolation.

\section{Removal of Background in White Light Total Brightness Observations}
\label{whitelight}

The coronal images of figure \ref{f2} are made from polarized brightness observations (at heights above 1.3$R_\odot$). In general, only a few pB observational sequences per day are made by LASCO C2, while total brightness white light observations are made far more frequently and at higher spatial resolution. It is useful therefore to make images of the quality of figure \ref{f2} directly from these white light images, which allows the detailed viewing of transient events and the creation of high time resolution movies. To achieve this, it is necessary to calculate a suitable background for subtracting from the total brightness image. The total brightness includes contributions from the F corona, stray light and other instrumental contributions which need to be removed as effectively as possible to properly reveal the electron corona contribution. In this section, we describe a novel procedure for creating such images.  

Processing of the total brightness data involves creating a background image of the corona which is subtracted from the desired image. The background image we use is a long term average of the unpolarized component of the C2 data. An image of the unpolarized white light component can be made by simply subtracting a calibrated LASCO C2 polarized brightness image from a calibrated total brightness image made close in time. This leaves us with an image of the unpolarized brightness contribution to the white light image. For stability, we make several such images over a period of many days (usually a whole rotation). We can than average these images to create the unpolarized background. Each $pB$ and corresponding total brightness image need to be examined for defects. Large and bright transient events will contaminate the background image and should be avoided. 

\begin{figure}
\centering
\includegraphics[width=0.9\textwidth]{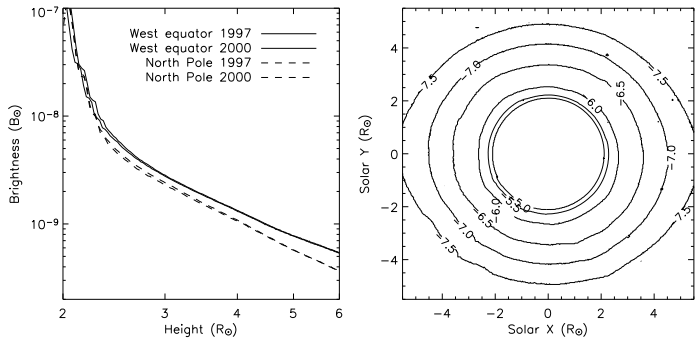}
		\caption{\textit{Left} - Unpolarized brightness plotted as a function of height above the North pole and West equator for 1997 January and 2000 December. This plot shows how the unpolarized background of LASCO C2 remains virtually unchanged between solar minimum and maximum. \textit{Right} - contours of log$_{10}$ unpolarized brightness background, formed by subtracting many pB images from total brightness images over the course of a solar rotation, and then averaging. }
\label{f3}
\end{figure}

The unpolarized background was found to change little with time. This is shown in the left plot of figure \ref{f3}, where unpolarized background brightness is plotted as a function of height above the North pole and West equator for 1997 January and 2000 December. Despite the large change in coronal topology between these times (see figure \ref{f2} for example), the unpolarized brightness background images are virtually identical. The unpolarized brightness has a steeper decrease over the poles than at the equator. The right contour plot of figure \ref{f3} maps the log$_{10}$ of unpolarized brightness in the field of view of LASCO C2.

Once an unpolarized background image is created, image processing is simple and quick. The LASCO C2 total brightness image of interest is calibrated and the unpolarized background image is subtracted from it. The corrected image can then be processed using the NRGF, resulting in images of equal or higher quality to those shown in figure \ref{f2}.

\begin{figure}
\centering
\includegraphics[width=0.94\textwidth]{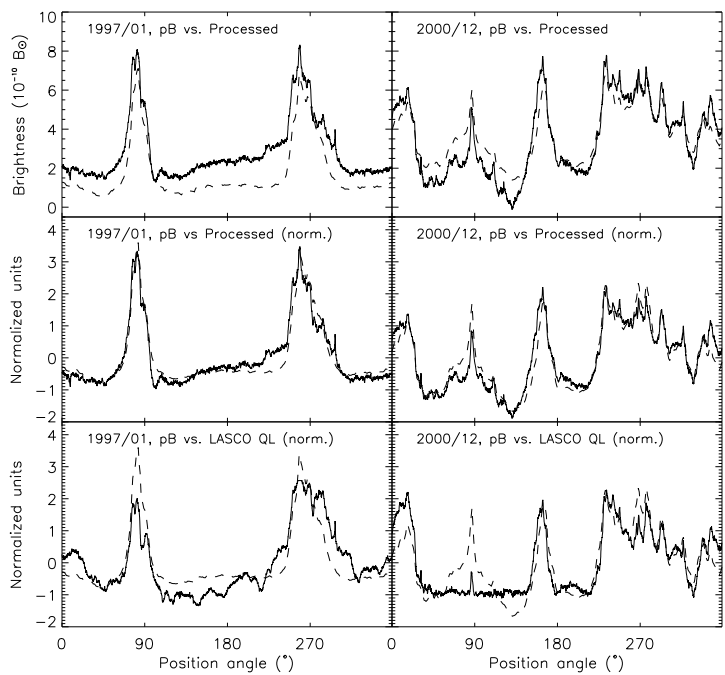}
		\caption{Latitudinal profiles at a height of 3.5$R_\odot$ comparing pB (dashed line) with total brightness (solid line) processed in two ways - with the new procedures outlined in this section and with the LASCO quick look software. The solar minimum (left column) pB observation was made on 1997/01/18 17:08 and the corresponding total brightness observation just over half an hour earlier. For solar maximum (right column) the pB observation was made on 2000/12/02 21:05 and the total brightness observation just over 20 minutes later. The top row shows pB and total brightness with the unpolarized background image subtracted. The middle row shows the same, but with both sets of profiles normalized to a mean of zero and a standard deviation of one to enable a fair comparison of structure. The bottom row shows pB and a profile processed with the standard LASCO quick look software. These are again normalized for ease of comparison.}
\label{f4}
\end{figure}

Confidence in the accuracy of coronal structure seen in the processed images can be established by comparison with a polarized brightness image. This is a simple approach which shows clearly if artifacts are introduced by the background subtraction, or if coronal structure is distorted in the processed image. In other words, if we process white light total brightness coronal images in order to view the underlying structure, that structure is expected to show general qualitative agreement with that seen in pB observations. In figure \ref{f4}, we make such comparisons for observations made during both solar minimum and maximum. Figure \ref{f4} (top row) shows a direct comparison of polarized brightness and total brightness with the unpolarized background subtracted. The differences between the two profiles show that the background-subtracted total brightness images are not suitable for a rigorous quantitative analysis. However, the middle row shows the same profiles of pB and background-subtracted total brightness both normalized to a mean of zero and a standard deviation of one. The profiles are very similar. Therefore the structure of the corona seen in a total brightness image processed using the new procedures of this section will be very similar to the structure we would see in a pB image. 

The bottom row of figure \ref{f4} compares pB with profiles obtained from images processed using the standard LASCO quick look software included in the Solar Software package. These images are widely used by the community to describe coronal structure. The quick look image processing reduces the F corona/stray light background contribution and decreases the radial gradient in brightness by dividing a total brightness image by a long term (monthly or yearly) minimum image. As seen in figure \ref{f4}, the quick look images often contain many artifacts and rather drastic distortions of coronal structure. For example, the pB profile shows that the peak brightness of the solar minimum equatorial streamers is almost the same for both east and west. The quick look image shows a far brighter streamer in the west. The west streamer also appears far broader than it should, and the east streamer shows a rather strange dip, or a secondary peak on its southern flank. The solar maximum streamer at 90$^\circ$ does not make an appearance in the quick look image at this height.

\section{The NRGF and Transient Events}
\label{cme}

\begin{figure}
\centering
\includegraphics[width=0.72\textwidth]{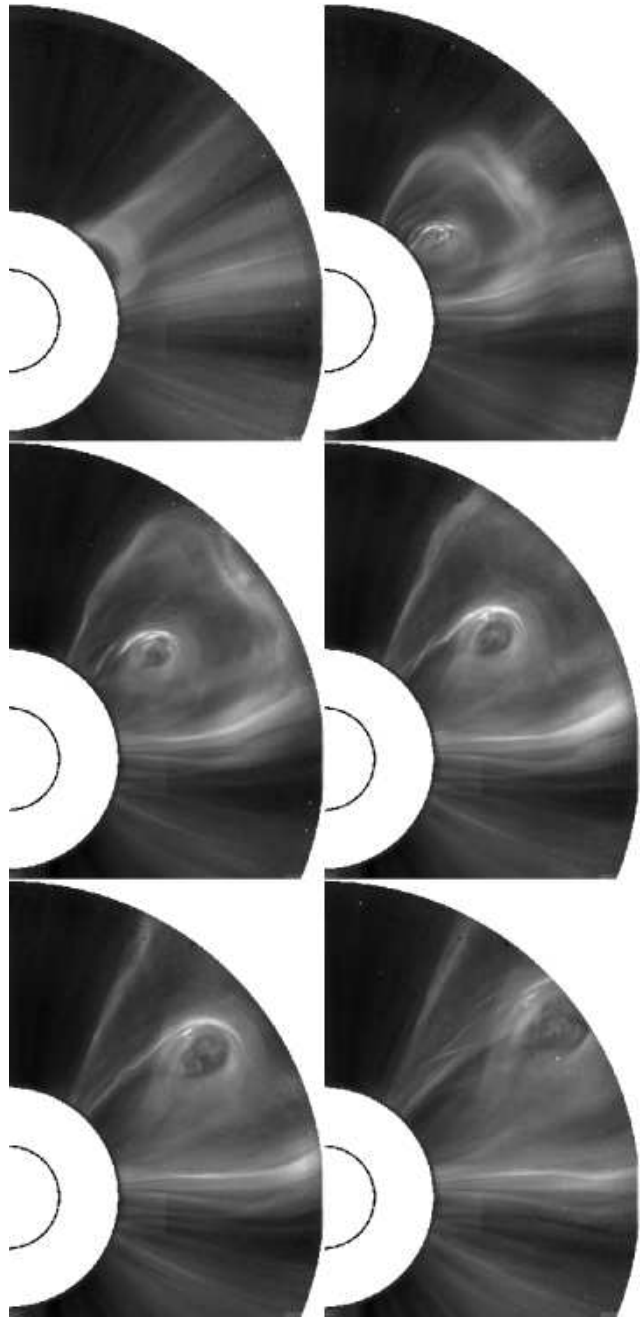}
\caption{ LASCO C2 observations of the North West corona made over the course of $\sim$6 hours starting on 2001 January 7 02:00, processed using the procedures of section \ref{whitelight} and the NRGF. The time sequence is from left to right, then top to bottom. The position of the Sun is shown as the black innermost circle. The displayed field of view is 2.2 to 6.1$R_\odot$.} 
\label{cmefig}
\end{figure}

The power of the NRGF and unpolarized background subtraction is demonstrated here by its application to a series of LASCO C2 WL observations of a CME propagating through the C2 field of view over the course of $\sim$6 hours starting on 2001 January 7 02:00. Figure \ref{cmefig} shows a selection of six images of the North Western corona, processed using the procedures of section \ref{whitelight}. For stability during the application of the NRGF, the average and standard deviation of brightness as functions of height are averaged from observations prior to and after the passage of the CME through the LASCO C2 field of view.

The CME can be seen in great detail. The most striking features are the relatively small dark central cavity surrounded by bright narrow filaments apparent in all images except the first, the large heart shaped filamentary outer loop of the ejection, seen best in the second and third images, and the faint twists of the filamentary structures following in the wake of the CME central cavity. The NRGF allows us to see the structure in more detail at all heights simultaneously without resorting to time difference imaging or edge enhancement image processing.

\section{Conclusions}
\label{conclusions}

Many fundamental questions arise regarding coronal structure as seen in white light images. In general, streamers often appear in the standard LASCO quick look images as solid blocks of brightness on a relatively flat background, suggestive of distinct homogeneous structures with clear boundaries. On the other hand, images processed using the NRGF show a myriad of filaments in all regions, and it becomes harder to define an edge between streamers and surrounding regions, even at solar minimum, as shown in figure \ref{f2} for example.  The image does contain bright structures extending throughout the corona, but regions in between these contain other, less bright, structures. Furthermore, the bright structures are not homogeneous, suggesting a deeper level of structure obscured by the line of sight and possibly smaller than the image resolution. Therefore, where do we draw the line between streamers and other regions? Are streamers simply regions of the corona where there are a higher number of filaments than surrounding regions? 

Given the importance white light coronal images have played, and continue to play in defining our impressions of the magnetic field measurements, it behooves us to use these images with caution. This paper shows how many aspects of structures in white light images produced with different image processing tools may be artifacts of the tools used. Images produced by the NRGF lend support to the results that have emerged from comparisons of latitudinal profiles of density as a function of radial distance \citep{woo1997a,hab2001} and the highly filamentary nature of the corona previously shown in radio measurements \citep{woo1997b} and white light images \citep{kou2004}. Given the future Solar Terrestrial Relations Observatory (STEREO) mission, which promises a wealth of white light coronal observations with high spatial and temporal resolution,  we anticipate that the NRGF will prove a valuable tool in the analysis and interpretation of these data. 

\begin{acknowledgements}
The pB data and coronal images used in this work are courtesy of the MLSO and the LASCO/SOHO consortium. The MLSO coronagraphs are operated by the High Altitude Observatory, a division of the National Center for Atmospheric Research, which is sponsored by the National Science Foundation (USA). SOHO is a project of international cooperation between ESA and NASA.
\end{acknowledgements}

\end{document}